\documentclass[aps,pr,reprint,groupedaddress]{revtex4-1}
\usepackage{floatflt}
\usepackage[final]{graphicx}
\usepackage{dcolumn}
\usepackage{graphicx}
\usepackage{amssymb}
\usepackage{epstopdf}
\usepackage[footnotesize]{caption}

\graphicspath{%
    {converted_graphics/}
    {/}
    {C:/personal/research/lowEscintx/}
}
\usepackage[compact]{titlesec}

\begin{document}

\title{Scintillation of Liquid Helium for Low-Energy Nuclear Recoils}

\author{T. M. Ito} 
\email{ito@lanl.gov}
\affiliation{Los Alamos National Laboratory, Los Alamos, New Mexico 87545, USA}

\author{G. M. Seidel}
\email{corresponding author; george\_seidel@brown.edu}
\affiliation{Department of Physics, Brown University, Providence, Rhode Island, 02912, USA}

\date{\today}

\begin{abstract}
The scintillation properties of liquid helium upon the recoil of a low energy helium atom are discussed in the context of the possible use of this medium as a detector of dark matter. It 1s found that the prompt scintillation yield in the range of recoil energies from a few keV to 100~keV is somewhat higher than that obtained by a linear extrapolation from the measured yield for an 5~MeV alpha particle. A comparison is made of both the scintillation yield and the charge separation by an electric field for nuclear recoils and for electrons stopped in helium. \\ 

\end{abstract}

\pacs{XXXXXXXXXXXX}

\maketitle

\section{Introduction}
\vspace{.1in}

The noble gas liquids have become a very attractive media for particle detection. Argon\cite{argon} and xenon\cite{xenon1,xenon2} are being used extensively in searches for dark matter particles. Neon is being investigated\cite{neon1,neon2} for possible application for detecting dark matter and neutrinos. And while helium has been proposed\cite{helium,helium2,helium3}, as well, as a target material for studies of neutrinos and dark matter, it has yet to be employed in such an application for a number of reasons.
But because of the possibility that dark matter may consist of WIMPs of lower mass than previously expected, in the range of 10~GeV or below, there is some reason to consider detectors that could provide better sensitivity in this low mass range. In that regard, it is natural to consider the advantages that liquid helium might provide over the heavier liquefied noble gases currently being used. To this end, this paper discusses the expected scintillation properties of liquid helium in the energy region below 100~keV where nuclear recoils resulting from elastic scattering of WIMPS would be expected to occur.\\

The recoil energy of a nucleus from which a WIMP is elastically scattered is
\begin{equation}
 E_{nr}=\frac{2\ m^2_x\ m_n\ v^2}{(m_x+m_n)^2}cos^2\theta \ ,
\end{equation}
where $m_x$ and $m_n$ is the mass of the WIMP and recoil nucleus, respectively, $v$ is the velocity of the WIMP with respect to the detector, and $\theta$ is the angle of the recoil nucleus relative to the direction of the motion of the WIMP. 
For a WIMP of mass 10~GeV in the galactic halo with a velocity of 250~km/s with respect to the solar system, the recoil energy of a helium nucleus is approximately 10~keV for $\theta =0$. Since dark matter particles are expected to have a velocity distribution up to the escape velocity from the galaxy of 680~km/s, the recoil energy of a helium nucleus will extend up to the 100~keV for 10~GeV WIMPs. \\

The heavy liquefied noble gases, argon and xenon, are being used as dark matter detectors since they possess a number of desirable properties. As cryogenic liquids they can be made with very high purity, they are reasonably dense so self shielding of background radiation is possible, and they have high scintillation yields being transparent to their own emission.
Furthermore, an electric field can be used to separate electrons from positive ions produced by ionization. Charge collection provides another valuable channel for identifying the nature of the radiation stopped in the liquid.\\

Liquid helium possesses some of these attributes of the heavier noble gas liquids but suffers as a potential dark matter detector for a number of reasons. 1) Scintillation from helium occurs at a higher energy, consisting of a broad distribution peaking at 16~eV. No material except for helium itself is transparent at this energy, requiring for detection of scintillation either the use of a wavelength shifter or the placement of the EUV detector directly within the containment volume of the helium. Additionally, because of the larger W-value of helium than the heavier noble gases, fewer photons are emitted per unit of energy deposition. 2) The use of liquid helium requires operating at much lower temperatures with the attendant technical cryogenic complexities. 3) Because of its low density liquid helium provides poor self shielding of background radiation. 4) Since electrons form bubbles and positive ions form snowballs in liquid helium\cite{Wilks}, the mobility of charges is significantly different in helium compared as to that of heavier noble gas liquids.
Not withstanding the challenges presented by these features, the potential benefit that helium, because of its low mass, brings to a search for light-mass dark matter particles makes the study of its properties in this regard of some interest. Hence this discussion of the scintillation efficiency of helium to low-energy nuclear recoils. We are unaware of any measurements or estimates of the emission efficiency of helium for low energy incident particles and use what is known about helium-helium scattering  and the scintillation of liquid helium upon stopping energetic alpha particles to fill this gap.   \\

The knowledge of the effect of a helium atom with low recoil energy, from 1 to 100~keV, in liquid helium is sparse. There are two types of information that is useful in developing an understanding what happens should a WIMP scatter from a helium atom. 1) Measurements exist on the ionization produced by He-He scattering in gases at the relevant energies. From this data it is possible to make some estimates of the likely consequence of WIMP scattering in the liquid. 2) From what is known about processes that occur along the track of an energetic alpha particle stopped in liquid helium it is possible to evaluate the likely consequences of processes that lead to quenching at lower energies.\\

This paper is organized as follows. Section II is focussed on a discussion of the production of ionization and excitation as a consequence of recoil from a WIMP. In Section III we review what is known about the interactions that occur along the track of an ionizing particle in liquid helium that are important for an understanding of the scintillation.  Section IV contains a calculation of the expected scintillation efficiency of a helium recoil as compared to the scintillation from an electron of the same energy, while Section V discusses the scintillation from electrons and compares charge collection from electron events and nuclear recoils. Section VI summarizes the results and limitations of the calculations.\\

\section{Helium-helium scattering}
\vspace{.1in}

\begin{figure}[htb] 
  \centering
  \includegraphics[bb=8 187 605 603,width=3.2in,height=2.8in,keepaspectratio]{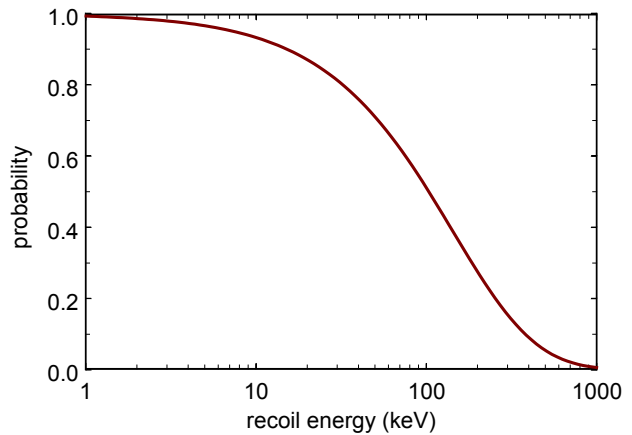}
    \caption{color online) Energy dependence of the probability that a helium recoil atom is in the neutral ground state, He$^0$, as calculated by Talman and Frolov\cite{Talman}.}
  \label{fig:He0creation}
\end{figure}

If a WIMP were to scatter off a helium nucleus and the recoil energy is low, the resulting recoil projectile would be expected to be the uncharged helium atom, He$^{0+}$. A calculation of the probability that the recoil atom is in such an un-ionized state is shown in Fig.~\ref{fig:He0creation}. However, if the recoil energy is high the recoil would likely 
be in a charged state, either the singly charged ion, He$^{1+}$, or the doubly charged, bare nucleus He$^{2+}$.  The crossover  from uncharged to charged state occurs on a scale determined by the atomic velocity, $v=e^2/\hbar = 2.19\times10^8\ \rm{cm\ s}^{-1}$. On average what happens subsequently to the projectile and the medium with which it interacts is independent of initial charged state, depending only on energy. A neutral atom can ionize target atoms or be stripped of an electron thereby being converted into a He$^{1+}$ ion. Numerous other processes such as charge exchange, electron capture, double ionization, ionization plus excitation, etc. can also occur. \\

\begin{figure}[htb] 
  \centering
  \includegraphics[bb=8 175 607 615,width=3.27in,height=3in,keepaspectratio]{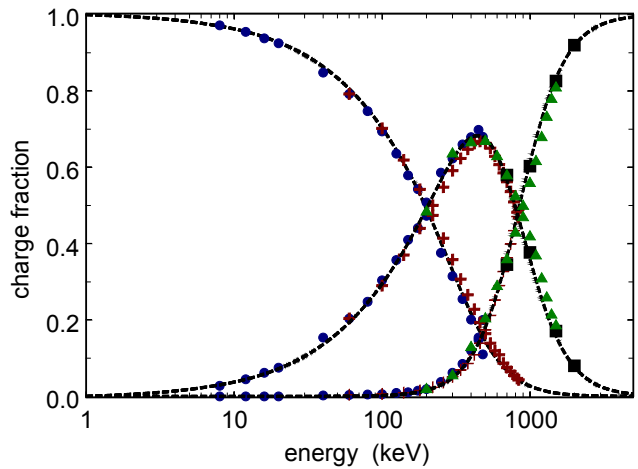}
  \caption{ (color online) Equilibrium charge fraction as a function of projectile energy for three states of helium. He$^{0+}$ is predominant at low energy and He$^{2+}$ at high energy. Circles Ref.\cite{Allison}; crosses Ref.\cite{Meckbach}; triangles Ref.\cite{Pivovar}; squares Ref.\cite{Itoh}.  Lines are empirical fits to the data.}
  \label{fig:chargefrac}
\end{figure}

The consequence of the energy dependence of various processes such as charge exchange and stripping is that the equilibrium probability of a projectile being in a particular charge state differs strongly with energy as illustrated in Fig.~\ref{fig:chargefrac}, where the experimentally measured equilibrium charge fraction for the three states is plotted as a function of energy. Since we are primarily concerned with recoil energies less than 100~keV, at which energy the charge fraction of He$^{2+}$, $F_2$, is less than 1\%, this charge state makes essentially no contribution to the expected scintillation signal from low mass WIMPs. Even the charge fraction $F_1$ for He$^{1+}$ is less than 30\% at 100~keV. \\ 
\vspace{.1in}

\vspace{.3in}

\subsection{Ionization and excitation}
\vspace{.1in}

\begin{figure}[htb] 
  \centering
  \includegraphics[bb=19 23 556 783,width=3.2in,height=4.53in,keepaspectratio]{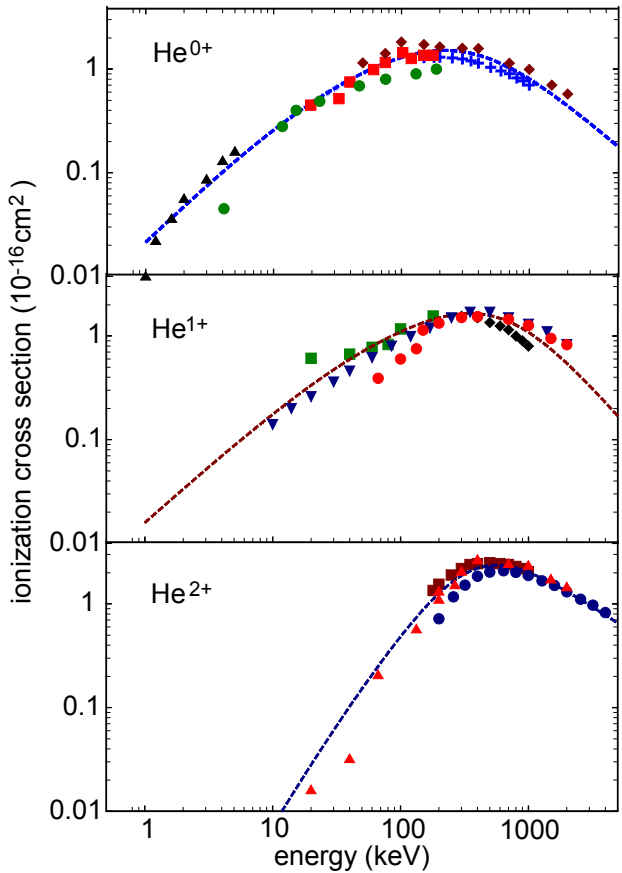}
    \caption{(color online) The energy dependence of the ionization cross section for the three charge states of a helium projectile incident on a helium target. Lines: empirical fits to the data. Data for He$^{0+}$: triangles Ref.\cite{Noda}, circles Ref.\cite{Barnett}, squares Ref.\cite{Solov'ev}, diamonds Ref.\cite{Sanders}, crosses Ref.\cite{Puckett}. Date for He$^{1+}$: triangles Ref.\cite{Rudd}, squares Ref.\cite{Solov'ev}, circles Ref.\cite{DuBois89}, diamonds Ref.\cite{Puckett}. Data for He$^{2+}$: triangles Ref.\cite{DuBois89}, circles Ref.\cite{Shah85}, squares Ref.\cite{Puckett}.}
  \label{fig:Heionall}
\end{figure}

\begin{figure}[htb] 
  \centering
  \includegraphics[bb=13 177 587 605,width=3.2in,height=2.38in,keepaspectratio]{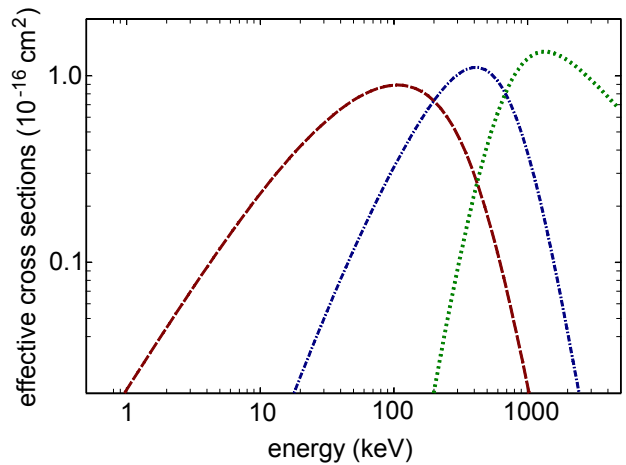}
    \caption{ (color online) Plots of the effective direct ionization cross sections, $F_i\sigma_{i,ion}$ for the three charge states of helium as function of energy. Lines: empirical fits; dashed line: He$^{0+}$; dot-dash line: He$^{1+}$; dotted line: He$^{2+}$.
 }
  \label{fig:sum-ion}
\end{figure}

There is a considerable body of information in the literature for the various processes that can occur when an energetic charged or neutral helium atom collides with another helium atom. The direct ionization process
\begin{equation}
 {\rm He}^{i+}+{\rm He}\rightarrow{\rm He}^{i+}+ {\rm He}^+ +e^- \ ,
\end{equation}
where a projectile in charge state $i$, which remains unchanged, ionizes a neutral target atom has been studied extensively. (The He with the superscript {\it i}+ denotes the projectile.) The experimentally measured cross sections for the three charge states are plotted in Fig. \ref{fig:Heionall}, and the effective ionization cross section for the various charge states of a helium projectile in helium is plotted in Fig.~\ref{fig:sum-ion}. The effective cross section is the ionization cross section multiplied by the respective charge fraction; that is, the plotted quantities are the products $F_i \sigma_{i,ion}$ for the three charge states. \\
\vspace{.2in}

\vspace{.3in}

\begin{figure}[htb] 
  \centering
  \includegraphics[bb=5 182 605 604,width=3.2in,height=2.25in,keepaspectratio]{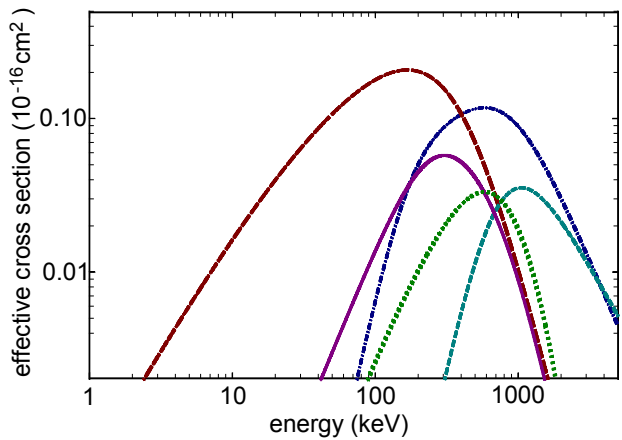}
   \caption{(color online) The effective ionization cross sections (cross section times charge fraction) for various double ionization and charge exchange processes. Long dashed line: simultaneous ionization of both target atom and He$^{0+}$ projectile, data from Ref.~\cite{Sanders}. Solid line: double ionization of target atom by He$^{1+}$, data from Ref.~\cite{Wood,Forest,DuBois89}. Short dashed line: double ionization of target by He$^{2+}$, data from Ref.~\cite{Shah85,Forest}. 
 Dot-dashed line: exchange $\sigma_{21}$, data from Ref.~\cite{Shah85,DuBois87}. Dotted line: exchange $\sigma_{20}$, data from Ref.~\cite{DuBois87}.}
  \label{fig:small}
\end{figure}

While processes in which the target atom is singly ionized dominate, double ionization can occur as well. The cross sections for double ionization are an order of magnitude smaller than for single ionization. The effective cross sections for the three charge states of the projectile are shown in Fig.~\ref{fig:small}.\\

In addition to direct ionization, ions are generated by processes in which the projectile changes its charge state -- called variously, exchange, capture, or stripping -- possibly accompanied by single or double ionization of a target atom. At low energies the most important of these processes, having a cross section normally labeled $\sigma_{10}$,  is one involving charge exchange in which a He$^{1+}$ projectile is neutralized, 
\begin{equation} 
{\rm He}^{1+}+{\rm He}\rightarrow{\rm He}^{0+} +{\rm He}^+\ .
\label{eq.s10}
\end{equation}
In charge equilibrium the rate at which this process occurs, $F_1 \sigma_{10}$, must be the same as the rate of the process where a neutral projectile is ionized
\begin{equation}
{\rm He}^{0+}+{\rm He}\rightarrow{\rm He}^{1+}+ {\rm He} +e^-\ , 
\label{eq.ion}
\end{equation}
namely, $F_0 \sigma_{0,ion}$, since viewed from the center of mass of the two interacting atoms there can be no distinction in Eq.~(\ref{eq.ion}) between which of the two atoms is ionized.  The two processes together results in a target atom being ionized and hence doubles the overall cross section for ionization in the energy range where He$^{0+}$ is the dominant charge equilibrium species.\\

Based on measurements in the literature we have plotted in Fig.~\ref{fig:small} the consequences of two other charge exchange processes. These make smaller contributions at higher energies to the overall ionization.  The effective ionization cross section $F_2 \sigma_{21}$ is associated with the process
\begin{equation}
{\rm He}^{2+}+{\rm He}\rightarrow{\rm He}^{1+}+ {\rm He}^{1+} \ ,
\label{eq.ex21}
\end{equation}
which in equilibrium must have the same rate as
\begin{equation}
{\rm He}^{1+}+{\rm He}\rightarrow{\rm He}^{2+}+ {\rm He} +e^- \ .
\label{eq.ex21b}
\end{equation}
Similar expressions describe the process related with $F_2 \sigma_{20}$ plotted in Fig.~\ref{fig:small}. \\

\begin{figure}[htb] 
  \centering
  \includegraphics[bb=5 193 587 603,width=3.2in,height=2.15in,keepaspectratio]{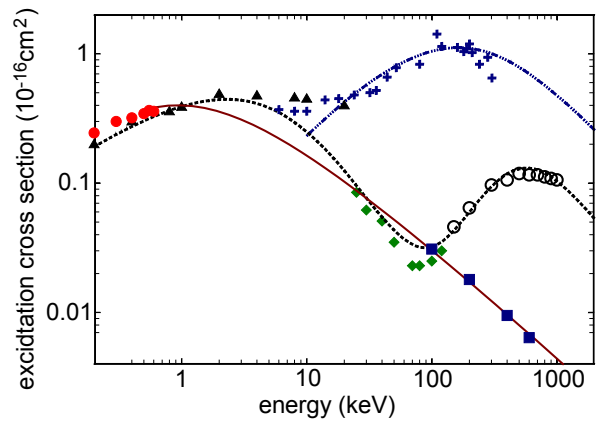}
    \caption{(color online) The cross sections for excitation of the 1$^1$S to 2$^1$P transition by He$^{0+}$ and by He$^{1}$ in helium. Data for He$^{0+}$: solid circles, Ref.~\cite{Kemptera} scaled to fit theory of  Refs.~\cite{Olson, Gauyacq}; squares, theory Ref.~\cite{Baynard}. Data for He$^{1+}$: triangles, Ref.~\cite{Okasaka}; diamonds Ref.~\cite{Pol}; open circles, Ref.~\cite{Hippler}.   Data for charge exchange plus simultaneous excitation by He$^{2+}$: crosses, Ref.~\cite{Folkerts}.  Lines are empirical fits to data and theory. }
  \label{fig:excit}
\end{figure}

 Scintillation also results from atoms that are promoted directly to excited states without having first been ionized. The cross sections for excitation of helium by helium are not as well studied as for ionization. The only measurement of the excitation cross section of He$^{0+}$ on He of which we are aware is that of Kempter {\it et al.}\cite{Kemptera}. Their results,  plotted for the specific transition 1$^1$S to 2$^1$P in Fig.~\ref{fig:excit}, are scaled as recommended by Kempter \cite{Kempterb} to agree with theoretical predictions \cite{Olson}.
The necessity for the scaling is the consequence of how the measurements were made, involving the detection of the UV radiation at 58.4~nm for the transition back to the ground state using a scintillator in conjunction with a photomultiplier whose calibration was not known to within a factor of 3.  In the low energy range below 10~keV, where the equilibrium charge state is predominantly 0+, excitations make a larger contribution to scintillation than ionization, as can be observed by comparing the excitation cross section for He$^{0+}$ in Fig.~\ref{fig:excit} with the cross section for ionization in Fig.~\ref{fig:Heionall}.\\

Although the transition from 1$^1$S to 2$^1$P state is by far the most likely excitation to occur, other transitions are non-negligible. Based on the measurements of Kempter\cite{Kemptera} and calculations by 
others\cite{Baynard, Olson, Gauyacq} for excitations created by helium in the 1+ state,  we estimate that for every transition having a 2$^1$P  final state there are 0.4 transitions to other states of which half are spin singlets and half triplets. This rough estimate results in a singlet to triplet ratio of .86/.14.\\ 

Experimental data for the excitation transition 1$^1$S to 2$^1$P by He in the 1+ charge state is also plotted in Fig.~\ref{fig:excit}. The two maxima for excitation by  result from different processes\cite{Okasaka}, the maximum at higher energy being related to the excitation of the target atom and the lower related to charge exchange and excitation of the projectile. We are unaware of any measurements of excitation of target He atoms by He$^{2+}$. The scaling dependence for excitation by charged projectiles at high energies given by $\sigma/Z =f(v^2/Z)$ \cite{Fritsch94}, suggests that it does not make a significant contribution to energy dissipation. Yet another process has been measured, which we do include in our analysis. Folkerts {\it et al.}\cite{Folkerts} have measured the combined cross section for a number of processes where the projectile is He$^{2+}$. These include the process of charge exchange, with either the resulting He$^{1+}$ projectile or target ion simultaneously promoted to an excited state, and the process of ionization and simultaneous excitation of the target. While these cross sections for excitation by He$^{1+}$ and He$^{2+}$ are comparable to that of  He$^{0+}$, they make little contribution to the scintillation yield because the respective charge fractions are small in the energy range where the cross sections are large. Again, in estimating the effects of these excitation processes we multiply the cross section of the 1$^1$S to 2$^1$P transition by 1.4 to approximate the total production of excitations.\\

There exist other processes, for example, a change in charge or excitation state of both projectile and target atom, but these have even smaller cross sections and we do not consider them further.\\

\subsection{Secondary Electrons}
\vspace{.1in}

One other mechanism is important for producing ions and excited-state atoms in helium, namely, the secondary electrons created by an initial ionization that have recoil energies greater than the ionization potential or the first excitation level. This process is unimportant at primary projectile energies below 100~keV, but it is a significant contributor at higher energies. An understanding of what happens at high energies is important in calculating quenching at low energies, so we discuss ionization by secondary electron here. \\

At high projectile energies, the energy distribution of secondary electrons is such that they can produce additional ionization. In a review Rudd {\it et al.}\cite{Rudd92} have given a semi-empirical expression for the single differential ionization cross section (SDCS) of secondary electrons, which depends on the projectile energy and the energy of the secondary electron. Their expression was generated from measurements of the differential cross section for the production of secondaries by protons stopped in helium. This expression can be modified to calculate the secondary distribution for a helium projectile by scaling the energy of the projectile by 4 to account for the difference in mass between a proton and helium and scaling the magnitude of the distribution by a factor of 4 (square of charge ratio) when computing the contribution of He$^{2+}$ compared to that of the proton.  Garibotti and Cravero\cite{Garibotti} have measured the SDCS for 4 and 7.36~MeV He$^{2+}$ ions in helium and find the scaled Rudd expression fits their data well but is slightly low at the higher electron recoil energies. \\ 

There exist accurate electron impact cross sections for helium for  both ionization\cite{Shah88,Rejoub} and excitation\cite{Kato} so that one can perform calculations, such as a Monte Carlo simulation, of the generation of ions and excited atoms starting from the energy distribution of recoil electrons. However, given the uncertainty in the energy distribution and the limited use to which the the results are to be put, we calculate the number of ionizations by determining the total energy in the secondary electron spectrum with energies above the ionization potential and dividing by the W-value of electrons in helium of 43 eV. The number of excitations is estimated by assigning 33~eV to each ionization, this number being the sum of 24.6~eV for the ionization and 8.4~eV for the average recoil energy of electrons below the excitation threshold. The remaining 10~eV is presumed associated with excitations, which have an estimated weighted average energy of 22~eV, yielding the number of excitations to ionizations of $10/22 =0.45$. This estimate is made for both the 1+ and 2+ charge states, multiplied by the respective charge fractions and converted into effective cross sections.\\

In the absence of any measurements of the secondary electron distribution from other ionization processes, the Rudd expression is assumed to be applicable to these cases as well, scaled for the appropriate charge state. An alternative procedure is to presume, as discussed by many authors, see for example, Manson and Toburen\cite{Manson81},  that for kinematic reasons the direction of the velocity of recoil electrons is sharply peaked in the forward direction and the magnitude of the velocity is peaked around that of the projectile. Hence, the ratio of the average recoil energy to the projectile energy is approximately $m_e/M_{He}$. The two approaches yield comparable results, given the approximations involved in the estimates.\\
\vspace{.2in}

\begin{figure}[htb] 
  \centering
  \includegraphics[bb=1 193 604 599,width=3.2in,height=2.65in,keepaspectratio]{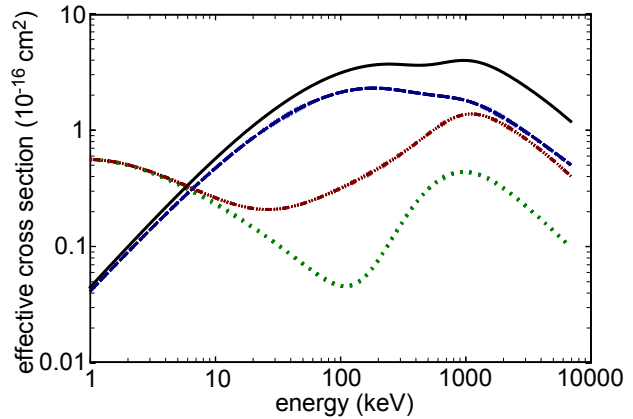}
   \caption{(color online) Sum of the effective ionization and excitation cross sections from Eq.~(\ref{eq.effcs}). Solid line: ionization with secondary electron contribution. Dashed line: ionization without secondary electron contribution. Dot-dashed line: excitation with secondary electron contribution. Dotted line: excitation without secondary electron contribution.}
  \label{fig:sum-eff-cs}
\end{figure}
\vspace{.3in}
The sum of the effective cross sections for ionization and excitation processes with and without the inclusion of the contribution of secondary electrons at high energy is plotted in Fig.~\ref{fig:sum-eff-cs}. The sums of the effective cross sections for ionization and excitation are
$$S_{ion,eff} = \sum_i\sum_j F_i \sigma_{ion,i,j} \,$$
and
\begin{equation}
S_{exc,eff} = \sum_i\sum_j F_i \sigma_{exc,i,j} \,
\label{eq.effcs}
\end{equation}
where the subscript $i$ refers to the three charge states and the subscript $j$ to the specific processes. One noteworthy feature of this plot is the very low effective excitation cross section in the energy range between 10 and 100~keV. While the excitation cross sections for He$^{1+}$ and He$^{2+}$ are of a size comparable to the corresponding ionization cross sections, the maxima of the cross sections for excitations occur at a lower energy where the charge fractions are considerably less.\\

\section{Stopping power}
\vspace{.1in}

The stopping power is the average energy loss of a projectile per unit path length due to all scattering processes occurring in the target material and is usually expressed in units of MeV~cm$^2$/g \cite{ASTAR} or, more conveniently for this discussion, in eV~cm$^2$\cite{Ziegler85}. The stopping power is normally divided into an electronic component, due to Coulomb interactions creating ionizations and excitations, and a nuclear component, the result of elastic collisions. The division is not without ambiguity, however, since it can be dependent on the time at which it is made \cite{Lindhard}. \\

To calculate the electronic stopping power from the compilation of effective ionization and excitations cross sections discussed in the previous section requires knowledge of the energy loss associated with each process. The electronic stopping power is
\begin{equation} 
SP = \sum_j S_{j,eff}Q_j \,
\label{eq.stop}
\end{equation}
where the sum is over all the processes involving ionization and excitation and $Q_j$ is the energy loss of the particular process labeled by the subscript $j$. The problem, then, to compute the stopping power is one of estimating the energy loss for each of the processes involved.\\

The energy loss for a simple excitation event is taken to be 22~eV on average. The energy loss for an ionization event initiated by either a He$^{1+}$ or He$^{2+}$ is the sum of the ionization 
energy plus the average kinetic energy of the recoil electron, which can be computed using the empirical expression of Rudd\cite{Rudd92}, discussed earlier. The electron recoil energy resulting from other ionization processes is similarly treated.  The energy loss of an interaction involving charge exchange requires the addition of the energy of the process by which the projectile returns to its initial state.\\

\begin{figure}[htb] 
  \centering
  \includegraphics[bb=2 177 596 600,width=3.2in,height=2.28in,keepaspectratio]{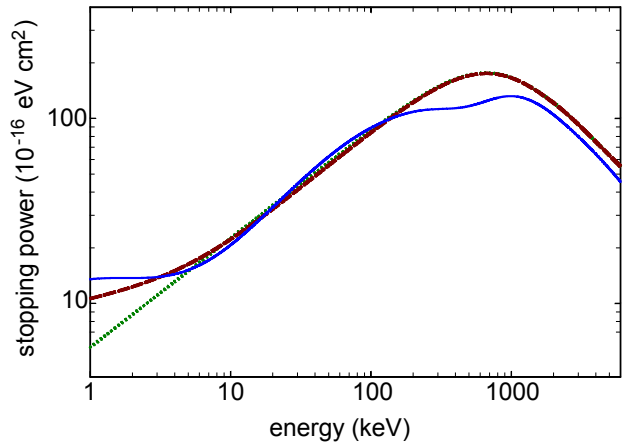}
   \caption{color online) The stopping power as function of energy. Solid line: calculated using Eq.~(\ref{eq.stop}). Dotted line: from ASTAR\cite{ASTAR}. Long dashed line: calculated from Lindhard theory\cite{Lindhard}.}
  \label{fig:stop}
\end{figure}

The result of computing the stopping power from Eq.~(\ref{eq.stop}), including all the microscopic processes discussed above, is shown in Fig.~\ref{fig:stop}. Also plotted in Fig.~\ref{fig:stop} is the electronic stopping power of an alpha particle in helium taken from the ASTAR tables\cite{ASTAR} converted from units of MeV~cm$^2$/g to $10^{-16}$~eV~cm$^2$ using the density of liquid helium. As 
illustrated in the graph, the two stopping powers are in reasonable agreement in some energy regions but differ considerably in others. Between 5~keV and 150~keV the two values are certainly within the accuracy of the calculations, but the difference of 50\% at 500~keV seems large even given the uncertainties associated with the various ionization and excitation processes that contribute at this energy, as shown in Fig.~\ref{fig:small}. Above 1~MeV the calculated stopping power is close to 20\% less than that measured, a difference which is hard to explain as being the result of uncertainties or approximations in the calculations. The only mechanism, discussed in the literature,  responsible for energy dissipation in this energy region is the direct ionization of the target helium by He$^{2+}$. Double ionization is more than a factor of $10^2$ less at 5~MeV. Direct excitation processes by He$^{2+}$ have not been reported in this region. As a means of bringing the calculated stopping power into agreement with the measured value, one might consider a very slight modification of the Rudd empirical expression for the single differential cross section for the recoil electrons. Such a modification would be consistent with the electron recoil data of Garibotti and Cravero\cite{Garibotti} as well, but it would have other consequences. As discussed below, the W-value calculated for alphas in helium using the effective cross sections is 38~eV at 5.5~MeV, well below the known value of 43~eV. Any modification of the secondary electron energy distribution to fit the stopping power makes the disagreement in the W-value larger. We are left to conclude that the absence of the direct excitation of target atoms by He$^{2+}$, not included in the measurements of Folkerts, {\it et al.}\cite{Folkerts} is missing. \\ 
\vspace{.2in}

\vspace{.3in}

 Below 5~keV the difference between the calculated stopping power and that given by the ASTAR tables has other origins. The electronic stopping power for alphas in helium has not been measured below 100~keV\cite{Ziegler}, and it is therefore not surprising that the tables at lower energies based on an empirical relationship between electronic stopping power and projectile energy of the form $SP\propto E^{.6}$ does not agree particularly well below 5~keV. The difference at 1~keV between the computed stopping power and the ASTAR values, larger than a factor of two, is also related to the fact that not all of the contributions to the stopping power calculated from Eq.~(\ref{eq.stop}) are included in the electronic stopping power from ASTAR. The total ASTAR stopping power consists of two components, electronic and nuclear, where the nuclear stopping power is the average rate of energy loss per unit path length due to the transfer of energy to recoiling atoms in elastic collisions. But neutral recoiling helium atoms can excite the electronic system as evidenced by the measured ionization and excitation cross sections for He$^{0+}$. This is the component missing from the ASTAR electronic stopping power.\\

The stopping power calculated using Eq.~(\ref{eq.stop}) also may miss the fraction of the energy that ends up as excitations at low energies. For example, should a He$^{0+}$ projectile elastically scatter giving the target atom sufficient energy to create an excitation, this would not be accounted for in  Eq.~(\ref{eq.stop}). However, the magnitude of this effect is not expected to be large.\\

The Lindhard theory \cite{Lindhard} of stopping power makes a different division of energy 
\begin{equation}  
E = \eta +\nu, 
\end{equation}
between nuclear, $\nu$, and electronic, $\eta$, components, a division that includes in the electronic term all the energy that ends up in ionization and excitation no matter if the origin scattering involved elastic collisions. At high projectile energies where elastic scattering is unimportant, $d\eta/dx$ is the same as the ASTAR electronic stopping power, but at low energies it is not since it contains ionizations and excitations produced by secondary recoiling He$^0$ atoms. At low energies Lindhard, on the basis of models for the various scattering cross sections, developed an analysis of the dependence of $\eta$ and $\nu$ on energy, from which it is possible to estimate how much ionization and excitation -- and, hence, the scintillation -- are enhanced over what would be calculated on the basis of the electronic scattering power alone. The Lindhard analysis\cite{Lindhard} develops a semi-empirical expression for $\nu$
$$ \nu = \frac{\epsilon}{1 + k\,g}$$
where $\epsilon = 11.5 E{\rm (in\ keV)} / Z^{7/3}$ is a reduced energy, the expression being valid when the charge of the projectile nucleus is the same as that of the target material. The constant is $k = .133 Z^{2/3} A^{-1/2}$, where $A$ is the nucleon number. The parameter $g$ is a function of the reduced energy, which is given graphically in Ref.~\cite{Lindhard} and in Ref.~\cite{Mei} by the empirical analytical expression $g  = 3 \epsilon^{.15} + .7 \epsilon^{.6} + \epsilon$.\\

From these considerations one can calculate $\eta$, the so-called nuclear quenching factor $f_n = \eta / E$ and $d\eta /dx$. These terms can be considered the Lindhard nuclear and electronic stopping power, respectively.  The quantity $d\eta /dx$ is plotted in Fig.~\ref{fig:stop}.  The agreement of the Lindhard theory with the results of Eq.~(\ref{eq.stop}) is quite good. The difference between the two curves at low energies, 25\% at 1~keV, could be due to approximations in the theory or an overestimate of the excitation stopping cross section for He$^{0+}$, which involves the use of a scaling factor\cite{Kemptera,Kempterb}.  \\

The agreement between the stopping calculated from a consideration of microscopic processes and obtained by other means, lends credence to our approach to estimating the numbers of ionizations and excitations produced by a low energy nuclear recoil in liquid helium.\\

\section{Interactions and scintillation in liquid helium}
\vspace{.1in}

The scintillation from helium not only depends on the mechanisms for production of ions and excitations; it also is influenced by nonradiative quenching processes that can occur in the liquid. For this we turn to a discussion of what is known about the difference in scintillation produced by energetic electrons and alpha particles in liquid helium.\\

In dense helium gas or in the liquid a helium ion He$^+$ or atom in an excited state He$^\ast$ will quickly combine with a ground state atom to form an excimer.
$$ {\rm He}^+ + {\rm He} \rightarrow {\rm He}_2^+ .$$ 
\begin{equation}{\rm He}^\ast + {\rm He} \rightarrow {\rm He}_2^\ast .
\end{equation} 
The excimer He$_2^*$ or He$_2^+$  has an inter nuclear distance of 0.12~nm and a binding energy of approximately 1.9 eV. 
In the liquid, positive ions form excimers quickly ($\sim$ 1 ps), while excited helium atoms generally take longer because of the extended radius of the outer electron.\\

The excimers formed on recombination rapidly radiatively cascade to the lowest excimer state, He$_2(a^3\Sigma_u)$ for triplets and He$_2(A^1\Sigma_u)$ for singlets. The radiative lifetime of the lowest singlet excimer to the dissociative ground state $(X^1\Sigma_g)$ is the order of $10^{-8}$~s and accounts for the prompt scintillation signal. While the energy of the  $(A^1\Sigma_u)$ state is roughly 20~eV above the well separated, ground-state helium pair, the fact that the transition satisfies the Franck-Condon principle and the energy of the ground state rises rapidly with decreasing inter nuclear separation accounts for the emission spectrum being a broad peak centered at 16~eV. The triplet state $(a^3\Sigma_u)$ has a measured lifetime of 13~s in liquid helium\cite{McKinsey99}, and its radiative decay does not contribute to the particle identification unless use is made of afterpulsing\cite{McKinsey03,Ito}.\\ 

Because of the long lifetime of the lowest lying triplet excimer,  the prompt scintillation signal depends on the ratio of the number of singlets to triplets produced by an ionizing particle. It also depends on the number and type of excited state atoms that are generated. The only estimates of the number of excitations produced for particles with energies above 100~keV, of which we are aware, are those of Sato {\it et al.}\cite{Sato76, Sato74} who calculate that 0.45 atoms are promoted to excited states for every ion produced. Of the excited atoms 85\% are predicted to be in spin singlet states and 15\% in triplet states.\\

However, on the basis of our present estimates of the effective cross section we are now able to estimate independently the ratio of excitations to ionizations, which can be found by a numerical integration of the plots in Fig.~\ref{fig:sum-eff-cs}. This ratio, obtained from a summation of the contributions from the various microscopic processes, 
has a magnitude and energy dependence that is somewhat different from that obtained theoretically by Sato {\it et al.}\cite{Sato76}, who did not consider the variation of the charge state of the projectile.\\

Using our value of 0.34 for the ratio of excitations to ionizations at 5.5~MeV and assuming the ratio of singlet to triplet excitations is .85/.15, we can estimate the fraction of energy that should appear as prompt scintillation. If the ions recombine in proportion to the number of available states, then 3/4 recombine in triplets  and 1/4 in singlets, and the fraction of deposited energy appearing as prompt scintillation should be
\begin{equation} 
\frac{16}{43}\times(1/4 + .34\times .85)= .20\ ,
\label{eq.ratio}
\end{equation}
where the first term arises from singlet excimers created on recombination and the second from singlet excitations. Calorimetric measurements of the scintillation from 5.5~MeV alphas indicate that, instead of 20\%, only 10\% of the energy appears as photons\cite{Bandler,Adams}, in contrast to measurements of 364~keV electrons for which the prompt scintillation yield is 35\%\cite{Bandler,Adams}. \\

The origin of the difference in scintillation yield for alpha particles and electrons lies in the very different stopping power of helium for electrons and alphas. For a 5.5~MeV alpha an ionization occurs on average every few nanometers along the track, but for an energetic electron the separation between ionizations is the order of 1000~nm. The mean distance a secondary electron with energy below the excitation threshold of 19.8~eV travels by diffusion before becoming localized by forming a bubble is estimated to be approximately 60~nm\cite{Ito,Guo}. Consequently, for ionization by electrons the recombination is primarily geminate, and the spins of the recombining ion and electron can be correlated. It is estimated from the 35\% scintillation yield that more than 50\% of the excimers formed on recombination in this case are singlets rather than the 25\% expected on the basis of number of available states.\\

Along an alpha particle track, the recombination is decidedly not geminate, and  the ratio of singlets to triplets should be 1 to 3. The decrease in scintillation yield by a factor of 2 from that calculated on the basis of this ratio is attributed to the nonradiative destruction of excimers by the exothermic Penning process, 
$$ {\rm He}_2^* + {\rm He}_2^* \rightarrow 3\ {\rm He}(1^1S) + {\rm He}^+ +e^-,$$
\begin{equation} 
{\rm or}\ \ \ \ \ \ \ \ \ \  \rightarrow 2\ {\rm He}(1^1S) + {\rm He}_2^+ +e^-.
\label{eq.penning}
\end{equation}

In either case, two excimers are destroyed and a new one is formed upon the recombination of the electron and ion. Keto {\it et al.}\cite{Keto,Keto2} were the first to measure the rate coefficient for this bimolecular process, $dn/dt = - \alpha\  n^2(t)$, for triplet excimers in liquid helium. These measurements have been repeated and extended by others\cite{Eltsov,Eltsov2}, but no direct observation of this Penning process has been observed for singlets. Nonetheless, it is presumed to be the cause of the quenching of the scintillation from a highly ionizing particle in liquid helium. \\

The same type of exothermic process illustrated in Eq.~(\ref{eq.penning}) can occur if one or both of the interacting species are not excimers but atoms in excited states, which may not have formed excimers prior to encountering another excited species. \\

This quenching of the scintillation signal, observed for energetic alpha particles, will also occur for low energy scattering by WIMPS if the density of excimers and excited atoms along the recoil track is comparable to that for an alpha.
Ito {\it et al.}\cite{Ito} have  made a rough estimate of quenching and its dependence on density of singlet atoms and excimers along an alpha track, and we use that approach to predict what is likely to occur for a low energy recoil. \\

The absence of knowledge about possible differences in the rates at which bimolecular Penning processes occur among the different excimers and excited state atoms, makes any rigorous calculation of electronic quenching impossible. Instead, we lump all the species together into a single differential equation for the rate of change in the total density, $n$, of all the species.
\begin{equation}
\frac{dn}{dt} = -\gamma n^2 - \frac{r\, n}{\tau}.
\label{eq:rate}
\end{equation}
The bimolecular rate $\gamma$ is taken to be the same for all interacting pairs while the radiative decay governed by the time constant $\tau$ is restricted to the singlet species by setting the constant $r$ to the value $r = (1/4 + .34\times .85)/(1+.34)= .40$ . Since we are only interested in quenching of the prompt signal with time constant of $10^{-8}$~s,  diffusion of excimers and excited atoms out of the dense cloud about the primary track can be neglected. The highly simplified Eq.~(\ref{eq:rate}) is useful for demonstrating the dependence of quenching on concentration. The quenching factor $f$ is defined as  
\begin{equation}
f = \frac{1}{n_0}\int_0^\infty \frac{r\, n}{\tau} dt = \frac{ln(1+\xi)}{\xi}\ .
\label{eq.factor}
\end{equation}
It is the fraction of singlet species that radiatively decay rather than are destroyed by bimolecular interactions and is related to $n_0$ through $\xi = n_0 \gamma \tau /r$. The  value for the quenching factor as determined calorimetrically for a 5.5~MeV alpha particle is $f= .10/.20 = .50$ (ratio of measured scintillation to predicted value\cite{Adams}), resulting in $\xi = 2.3$. Because of the different ratio of excitations to ions used here compared to that we used previously\cite{Ito}, the value of $f$ also differs.\\

The use of Eq.~(\ref{eq:rate}) to estimate the effect of quenching through Eq.~(\ref{eq.factor}) involves another  simplification necessitated by a lack of information at the microscopic level. It assumes that the density of the interacting species is uniform both along the track and perpendicular to it, which is certainly not the case. It is relatively straightforward to account for the density variation along the track by assuming the number of excimers and excited atoms produced is proportional to the stopping power (or stopping cross section). The effect of allowing for this variation along the track is to add a multiplicative term in the relationship between $\xi$ and $n_0$ or, alternatively, to change the value of the rate coefficient, $\gamma$, by a corresponding amount. For the case of a 5.5~MeV alpha the change is only about 15\%, and we do not consider it further. The variation in density perpendicular to the track is not so easily treated. It is possible that the spatial distribution of excimers formed on the subsequent recombination of ions and electrons may differ from that for excited state atoms or excimers formed from them. All we can assume is that the distribution is the same, independent of the energy or the primary particle and that relationship between quenching factor and density discussed above remains valid in comparing different energy depositions.\\ 

\section{Calculation of scintillation yield}
\vspace{.1in}

\begin{figure}[b] 
  \centering
  \includegraphics[bb=10 191 605 601,width=3.2in,height=2.2in,keepaspectratio]{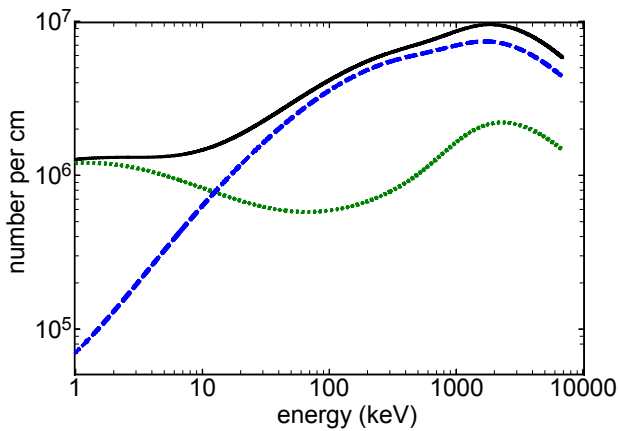}
   \caption{(color online) Average number of ions and excited atoms per cm along track of projectile. Dotted line: excitations. Dashed line: ions. Solid line: sum of the two species. }
  \label{fig:track}
\end{figure}

\begin{figure}[tb] 
  \centering
  \includegraphics[bb=4 190 604 601,width=3.2in,height=2.19in,keepaspectratio]{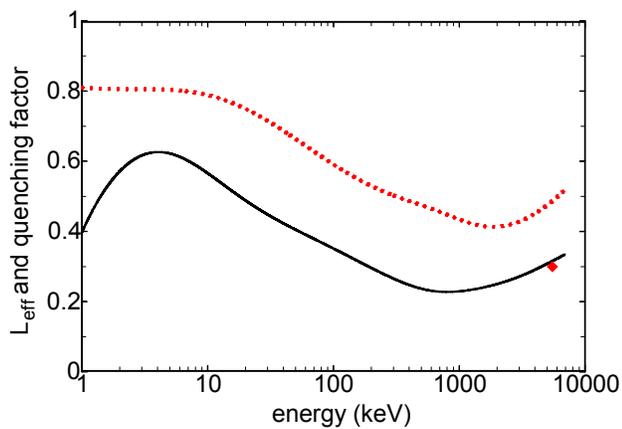}
   \caption{(color online) Solid line: relative scintillation efficiency, ${\mathcal L}_{eff}$ as a function of He projectile energy. Dotted line: quenching factor due to bimolecular processes.}
  \label{fig:Leff}
\end{figure}

The average number of ions and excited atoms per unit length deposited along the track as a function of the initial energy of a helium projectile is plotted in Fig.~\ref{fig:track}. The number per unit length of all species that can partake in the Penning process changes significantly with energy, varying from $6.8\times 10^6$ per cm at 5.5~MeV where the electron quenching is measured to be 0.50 to $1.5 \times 10^6$ per cm at 10~keV. At this latter average number per cm the quenching factor, using Eq.~(\ref{eq.factor}) and the assumption of the same spatial distribution of particles perpendicular to the track, is calculated to be 0.80, so that there is still a 20\% reduction in scintillation for nuclear recoils resulting from nonradiative decays.  The quenching factor for prompt scintillation is plotted in Fig.~\ref{fig:Leff}.\\

\begin{figure}[t] 
  \centering
  \includegraphics[bb=6 189 597 601,width=3.2in,height=2.23in,keepaspectratio]{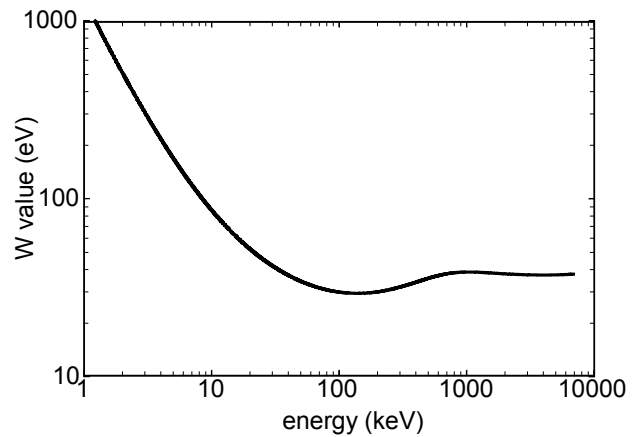}
  \caption{The calculated W-value for a helium projectile in liquid helium. }
  \label{fig:W-value}
\end{figure}

One consequence of the variation in number of ionization and excitations with energy of the primary particle is the W-value does not remain a constant with energy as shown in Fig.~\ref{fig:W-value}. At 5.5~MeV the calculated W-value is 38~eV, below the known value, increases slightly at 1~MeV and then drops to a minimum of 30~eV at 100~keV.  But at 5~keV it is has a value of 160~eV. This rise in the W-value at low projectile energy is the result of the decrease in probability of ionization compared to excitation of helium atoms in this low energy range. The trend of increasing W-value for helium recoils with decreasing energy is what is expected theoretically\cite{Sato76}, and has been observed to occur in a number of  pure gases and mixtures\cite{Jesse,Tawara}. However, we can find no report of a  measurement of this phenomenon in helium. \\

The W-value for electrons in helium remains essentially constant above 1~keV, since the W-value is insensitive to the energy of the projectile as long as its velocity is much higher than that of the valence electrons\cite{Douthat,Inokuti}. \\
 
While the dependence of the scintillation yield on recoil energy is affected by electronic quenching, it is also strongly influenced by the change in ratio of excitations to ionizations at low energies. Ionization are expected to produce singlets to triplets in the ratio of 1 to 3, whereas excitations, as discussed above in considering their creation by  stopping of He$^{0+}$ are presumed to create singlets far more copiously, the ratio of singlets to triplets being 0.86 to 0.14. \\

\begin{figure}[tb] 
  \centering
  \includegraphics[bb=1 182 611 601,width=3.2in,height=2.2in,keepaspectratio]{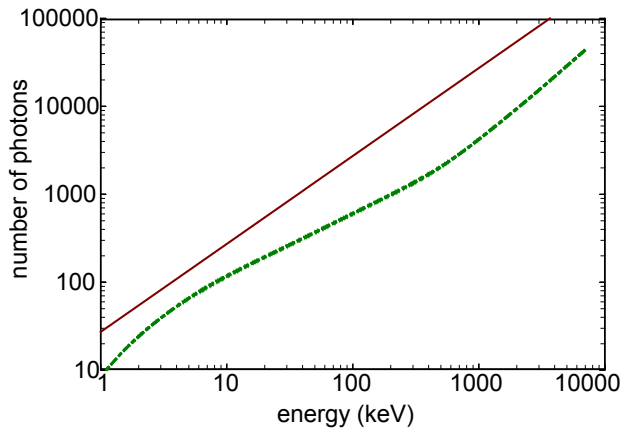}
   \caption{(color online) Number of prompt UV photons produced by primary particle stopped in liquid helium. Dashed line: helium atom recoil/alpha particle. Solid line: electron.}
  \label{fig:scint}
\end{figure}

The number of prompt UV photons from singlet excimers and excited state atoms is plotted as a function of the recoil energy helium atom in Fig.~\ref{fig:scint}. The data for this graph was obtained by summing the effective cross sections discussed in Section II, adding the contribution of secondary electrons to account for behavior above 100~keV, and correcting for electronic quenching. The production of excitations by secondary electrons was incorporated following Sato\cite{Sato74}.\\

The expected UV scintillation when electrons are the primary particles stopped in helium is also plotted in Fig.~\ref{fig:scint}. Since the ionization density along the track of an electron is so low the recombination is geminate. No quenching occurs. The G-values for ionization and for excitation of helium by electrons is independent of energy above 1~keV\cite{Douthat,Sato74}, so the scintillation yield is expected to be linear in electron energy. As discussed above, Adams\cite{Adams} has measured that 35\% of the initial kinetic energy of an electron stopped in liquid helium appears as photons. The corresponding number  for alphas at 5.5~MeV is 10\%. Hence the number of UV photons is 3.5 times larger for electrons than for alphas at high energies and remains larger, although by a lesser amount, in the range below 100~keV. The relative scintillation yield, $\mathcal{L}_{eff}$, is simply the ratio of the two curves in Fig.~\ref{fig:scint}\cite{foot}.  The energy dependence of ${\mathcal L}_{eff}$ is plotted in Fig.~\ref{fig:Leff}. \\

\section{Discrimination}
\vspace{.1in}

The utility of any medium as a dark matter detector is dependent on the ability to distinguish the signal produced by nuclear recoil from that of produced by background, principally electrons from beta decay and Compton scattering. Hence we discuss the difference in time dependence of the scintillation response to electrons  and nuclear recoils stopped in helium and the use of charge collection for discrimination.\\

\subsection{Charge collection}
\vspace{.1in}

\begin{figure}[tb] 
  \centering
  \includegraphics[bb=4 166 611 625,width=3.2in,height=2.42in,keepaspectratio]{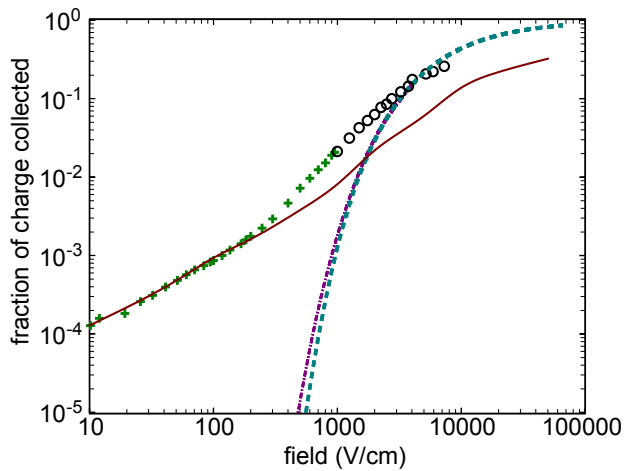}
   \caption{(color online) Fraction of ion/electron pairs that are separated and do not recombine as a function of applied electric field in liquid helium. Solid line: $\alpha$ particles, from Ref.~\cite{Williams}. Crosses: electrons, from Ref.~\cite{Ghosh}. Open circles: electrons, from Ref.~\cite{Seth}. Dashed line: electron distribution given by Eq.~ (\ref{eq.sphere}) with $\xi =56$~nm, separation constrained as in Eq.~(\ref{eq.recomb}) and no diffusion. Dot/dashed line: Monte Carlo simulation including diffusion.}
  \label{fig:charge}
\end{figure}

Ito {\it et al.}\cite{Ito} have measured the field dependence of the scintillation from alpha particles in helium to fields up to 45~kV/cm. At this field the scintillation is decreased by 15\% from  the zero field value. They used Kramers theory\cite{Kramers} of columnar recombination to fit the field dependence of the ionization current generated by alphas stopped in liquid helium as measured by Gerritsen\cite{Gerritsen}. A cylindrical Gaussian distribution of the initial charges about the track of the form, 
\begin{equation}
n_0(r) = \frac{N_0}{\pi b^2}r e^{-r^2/b^2}\ , 
\label{eq.cylind}
\end{equation}
where $N_0$ is the total number of ionizations, produced a reasonable fit with $b = 60$~nm. However, this distribution does not provide a good approximation to the ionization current for an alpha particle at low applied fields as measured by Williams and Stacey\cite{Williams}. Their data is reproduced in Fig.~\ref{fig:charge}, normalized to the total number of ionizations produced. We take this data to be a better measure of the charge separation with field expected for low energy He nuclear recoil. The lower initial charge density along the track of a low energy recoil as compared to that of an alpha is likely to increase the field dependence somewhat for low energy nuclear recoils, but without a more realistic model of the distribution of charge along the track any estimate of the change is unwarranted.\\

For electrons in helium, Guo {\it et al.}\cite{Guo} accounted for the variation of geminate recombination and the decrease in scintillation with electric field by fitting a spherical distribution in separation of an electron from the positive ion from which it originated by the expression 
\begin{equation}
n_0(r) = N_0\frac{32}{\pi^2 \xi^3}r^2 e^{-4r^2/\pi \xi^2}\ . 
\label{eq.sphere}
\end{equation}
In the absence of diffusion a pair will recombine in a field $E$ depending on the initial separation $r$ and orientation $\theta$ of the separation with respect to that of the field. When the condition 
\begin{equation}
 \frac{e}{4 \pi \epsilon_0 r^2} [1 + tan^2(\theta/2)] > E.
\label{eq.recomb}
\end{equation}
is valid, then the pair will recombine, otherwise they will not. The effect of diffusion can be accounted for numerically by performing a Monte Carlo simulation or, alternatively, by using the analytic expressions developed by Que and Rowlands\cite{Que} for the Onsager theory\cite{Onsager} of geminate recombination. Guo\cite{Guo} found that the change in scintillation at a field of 2700~V/cm could be fit with $\xi =56$~nm in Eq.~(\ref{eq.sphere}). A plot of the fraction of charge that would be extracted as a function of field for the distribution chosen by Guo is shown in Fig.~\ref{fig:charge} both with no diffusion and with diffusion appropriate for electron bubbles and positive ion snowballs in liquid helium at 2.5~K. Such a distribution fails to fit the measured charge collection as measured by Ghosh\cite{Ghosh} and Sethumadhavan\cite{Seth}.\\

Ghosh\cite{Ghosh} in experiments on electron bubbles in helium used a $^{63}$Ni (beta emitter with an end point of 66~keV) as a source. In the course of those experiments he measured the current created in a pair of electrodes as a function of field in the liquid. He also measured the current in helium gas so as to obtain the saturation current, that is, the complete charge separation of ionization events. His results obtained at 2.5~K for the liquid are plotted in Fig.~\ref{fig:charge}. He found a small dependence of current on temperature but not sufficient to warrant discussion here. Also, plotted is an extension of Ghosh's results to higher fields by Sethumadhavan\cite{Seth}. What is clear is that the model assumed by Guo does not predict charge separation properly at low fields. We do not have a theoretical understanding of the initial ion distribution produced by electrons that is adequate for explaining their subsequent separation by an applied field. It is coincidental that the charge separation for alpha particles and for electrons as illustrated in Fig.~\ref{fig:charge} is the same for fields less than 200~V/cm. If instead of using Ghosh's data taken at 2.5~K, the data at 4.2~K were plotted, the curves would differ by more than 50\%. What is presumably also coincidental is that the field dependence of the charge separation in the low field region below 200~V/cm is consistent with the expression 
\begin{equation}
Q/Q_0 = aE\,ln(1+1/aE)\ ,
\label{eq.thomas}
\end{equation}
which is the form of the field dependence derived by Thomas and Imel\cite{Thomas} with the constant $a = 10^{-5}$~cm/V.  Above 200~V/cm the expression given by Eq.~(\ref{eq.thomas}) bears no relation to the measurements.\\

\subsection{Afterpulsing}
\vspace{.1in}

 Scintillation resulting from metastable triplets excimers $(a^3\Sigma_u)$ with a lifetime of 13~s has been discussed\cite{Habicht,McKinsey03, Archibald}  as a means of discriminating between electron and nuclear recoils. The number of single photon events in the first few microseconds after the prompt signal with a $10^{-8}$ decay time depends on the density of ions and excitations along the track of the projectile. The delayed, discrete single-photon scintillation, called "afterpulsing", is not the result of the radiative decay of triplet excimers, a much too infrequent process to explain the observed rate, but is rather the consequence of the Penning annihilation of a pair of triplet excimers that results in the creation of a singlet that immediately radiatively decays\cite{King}. \\

A calculation of the magnitude of the afterpulsing and its time dependence for nuclear recoils and for electrons stopped in helium is complicated by the dependence of the Penning bimolecular process on distribution of interacting species about the track of the primary projectile and the diffusive motion that leads to their encounter. McKinsey {\it et al.} \cite{McKinsey03} showed at 1.8~K that the magnitude of afterpulsing normalized to the size of the prompt scintillation signal was five time greater for a 5.3~MeV alpha particle than for a 1~MeV electron. However, the highly ionizing products of the capture reaction $^3$He(n,p)$^3$H, with a combined recoil energy of 764~keV, produce only three times more afterpulsing than electrons. This variation of afterpulsing with energy cannot be explained without a more detailed knowledge than is currently available of the parameters and mechanisms affecting the process of afterpulsing. What happens at nuclear recoil energies below 100~keV is an open question.\\

Afterpulsing is also dependent on temperature of the liquid. For alphas the magnitude rapidly decays below 1~K as the quasiparticle density decreases and diffusion of the excimers away from the track is enhanced\cite{McKinsey03}. Also, an electric field decreases afterpulsing\cite{Ito}.\\

\section{Discussion}
\vspace{.1in}

The large increase in the relative scintillation efficiency, ${\mathcal L}_{eff}$, by more than a factor of two between 100~keV and 5~keV as illustrated in Fig.~\ref{fig:Leff}, qualitatively corresponds to the behavior of this quantity in the other liquefied noble gases. Both in neon\cite{Gas12} and in argon\cite{Regenfus} the relative scintillation efficiency increases with decreasing energy below 50 keV down to 10 keV. This increase is primarily the result of the  growth in excitation relative to ionization at low energy. Bezrukov {\it et al.}\cite{Bezrukov} recently predicted the relative scintillation efficiency for liquid xenon using the electronic and nuclear stopping powers together with an analysis of recombination. They, too, note an increase in ${\mathcal L}_{eff}$ with decreasing recoil energy.  There are no observations in the other noble gas liquids of the decrease in relative efficiency below 5 keV predicted for helium, as seen in Fig.~\ref{fig:Leff}. This decrease is associated with the increasing fraction of energy going into the nuclear recoil channel. An analysis similar that performed here on helium does not appear possible for the heavy noble gases given the absence of data on cross sections for both ionization and excitations by nuclear recoils. \\

This discussion of the scintillation yield of liquid helium for low energy nuclear recoils is based to a large degree on measurements of ionization, charge exchange and excitation processes by helium ions in various charge states. This approach is not without its problems. The measured cross sections have, in many cases, considerable uncertainty. The energy deposition associated with them is even less well known. Theory of these atomic collisions with many electrons is not of help except in certain cases at low or high energies. Nonetheless, the reasonable agreement between the energy dependence of the stopping cross section obtained from a summation of microscopic processes and that generated from the nuclear and electronic stopping powers suggests the approach has validity. As discussed earlier, the most prominent difference between the stopping power curves in Fig.~\ref{fig:stop}, occurring in the energy range from 200~keV to 1~MeV, is to be due to the improper estimate of the energy deposition of charge exchange processes. The low value of the calculated W-value and the stopping power above 1~MeV  is more likely the result of excitation processes that have not been accounted for. Fortunately, these deficiencies are not of serious concern in predicting the scintillation behavior for nuclear recoils below 100~keV from WIMPS.   \\

The overall agreement between the stopping power calculated as a sum of all the contributing processes and that from
ASTAR~\cite{ASTAR} is 20\% or better. From this and the consideration of the uncertainty in the singlet to triplet ratio, we estimate that the overall uncertainties of our calculated scintillation yield for low energy nuclear recoils is 30\%. At
energies below 10~keV, where excitations are the dominant component of the interacting species, the uncertainty may be larger due to the potential inaccuracy of the assumption that the radial spatial distribution of all interacting species is the same. The radial spatial distribution of excimers formed on recombination of ions and electrons is dictated by the diffusion of electrons~\cite{Ito,Guo}, whereas this is not the case for excitations. A different radial spatial distribution results in a different number density of interacting species, affecting the quenching factor from bimolecular processes and thus the scintillation yield.\\

The use of cross sections measured in the gas phase of helium, can be reasonably be assumed to be applicable to what happens in the liquid. Cooperative effects in the liquid occur in what happens along the track with bimolecular Penning processes and with charge separation in an applied electric field. These phenomena can depend upon density and diffusion and hence exhibit a temperature dependence.  \\

\begin{figure}[tb] 
  \centering
  \includegraphics[bb=49 156 652 406,width=3.2in,height=1.33in,keepaspectratio]{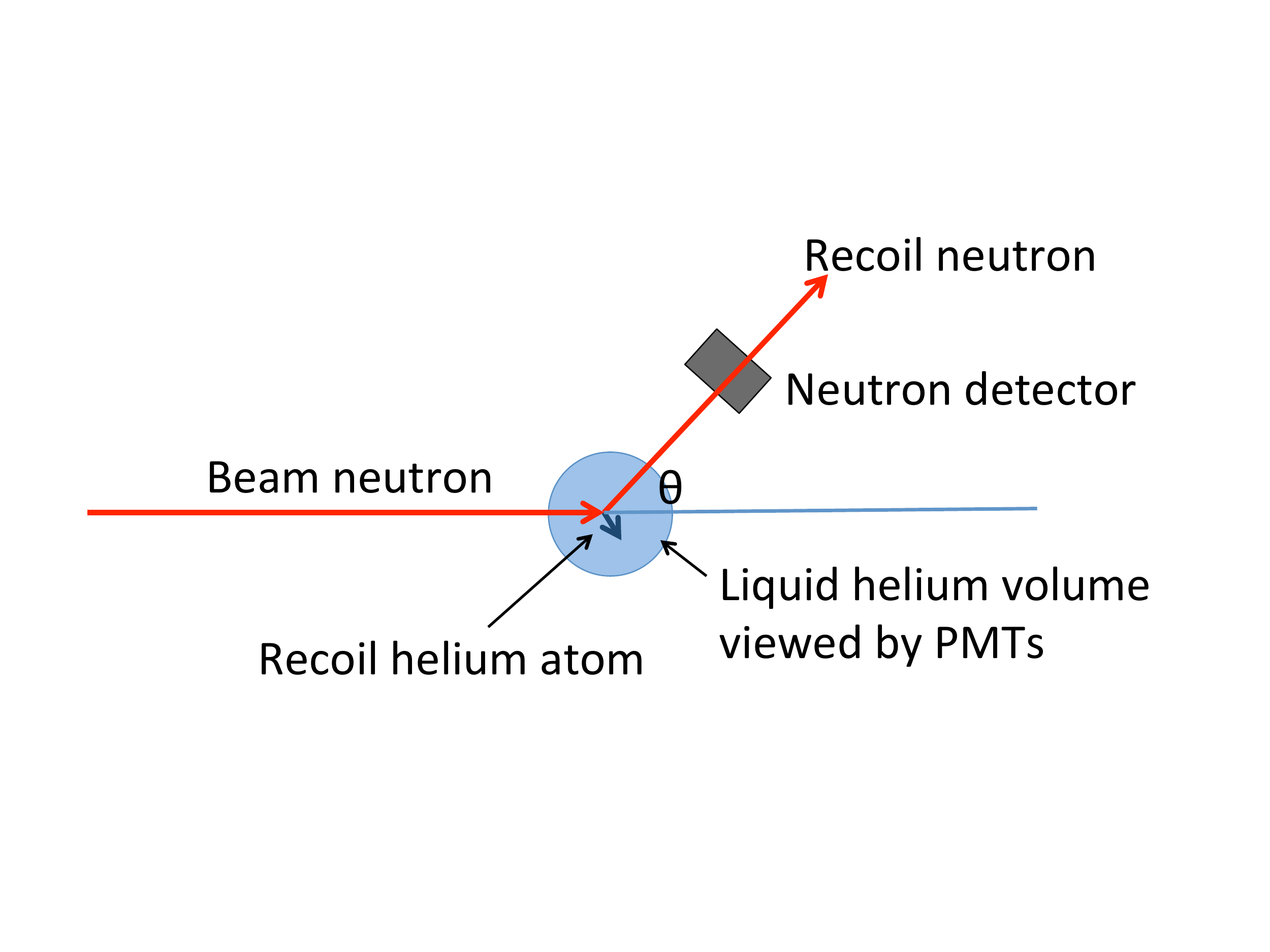}
 \caption{(color online) Layout of a possible experiment to measure the liquid helium scintillation
  yield for low-energy nuclear recoils.}
\label{fig:setup}
\end{figure}

Prior to any serious consideration of the use of liquid helium to detect low-energy nuclear recoils, it would be highly desirable to perform an experimental measurement of scintillation yield as a function of recoil energy. This can be achieved by introducing a neutron beam of known energy into liquid helium and detecting the scattered neutrons at a known angle (see Fig.~\ref{fig:setup}). In fact, the scintillation yield of heavier noble gas liquids (neon, argon, and xenon) have been studied in this manner using neutrons from a D-D generator (see e.g. Refs.~\cite{neon2,Man10,Pla11,Gas12}). Neutrons from a D-D generator, however, have a energy of $\sim 3$~MeV, too high for producing nuclear recoils with energies of a few keV to 100 keV in liquid helium. A feasibility study indicates that the WNR facility at the Los Alamos Neutron Science Center (LANSCE) at Los Alamos National Laboratory\cite{WNR} is suitable for this purpose. The WNR facility is a spallation neutron source driven by pulsed proton beams and provides neutrons with energies between 100 keV and a few 100 MeV. The neutrons energy can be determined on the event by event basis using the time of flight of neutrons between the time at which a proton pulse strikes the spallation target and the time at which a neutron detected by the detector. With a scattered neutron detector fixed at one angle, it is possible to map the scintillation efficiency
as a function of the recoil energy from a few keV all the way to the region where data exists based on alpha particle sources. \\

\section{Summary and conclusions}
\vspace{.1in}
This paper contains a discussion of the scintillation properties of liquid helium for low energy nuclear recoils in the context of the possible use of this medium as a dark matter detector. We first review the available cross section data on ionization and excitation of helium atoms due to collisions with helium atoms and ions. As a confirmation of the validity of our understanding of the ionization and excitation processes in liquid helium, the stopping power is calculated for a
helium atom or ion incident on helium as a sum of all the contributing microscopic processes. The resulting  calculated stopping power is in reasonable agreement with the widely used empirically determined stopping power. We then turn to what is known about scintillation processes the liquid helium generated by 5 MeV alpha particles. Nonradiative processes that quench the scintillation are also considered. Combining this information, we calculate the liquid helium scintillation efficiency for low energy nuclear recoils. The prompt scintillation yield thus obtained in the range of
recoil energies from a few keV to 100~keV is somewhat higher than that obtained by a linear extrapolation from the measured yield for an 5~MeV alpha particle. Furthermore, we compare both the scintillation yield and the charge separation by an electric field for nuclear recoils and for electrons stopped in helium. We also discuss a
possible experiment to test the results of our calculations.\\

\begin{acknowledgments}
We appreciate receiving a copy of a preprint of a paper by W. Guo an D.N. McKinsey\cite{Guo13} that covers some of the material discussed here. This work was supported by the US Department of Energy and the National Science Foundation.
\end{acknowledgments}


\begin{thebibliography}{99}
\bibitem{foot} There appear to be two different definitions of the relative scintillation yield in the
  literature. One is ${\mathcal L}_{eff}(E) =   Y_{nr}(E)/Y_{er}(122\;{\rm keV})$, where $Y_{nr}$ and $Y_{er}$
  represent the scintillation yields for nuclear recoils and electron   recoils, respectively. This definition is seen to be used for liquid   xenon~\cite{Man10,Pla11}. The other is ${\mathcal L}_{eff}(E) =   Y_{nr}(E)/Y_{er}(E)$, which is used for argon and  neon~\cite{neon2,Gas12}. We employ the latter definition in this paper.
\end{thebibliography}
\end{document}